\documentclass[preprint,aps,amsmath,amssymb]{revtex4}

\usepackage{graphicx}
\usepackage{dcolumn}
\usepackage{bm}
\begin{document}
\title{An open cavity formed with a photonic crystal of negative refraction}

\date{\today}
\author{Zhichao Ruan}
\author{Sailing He}

\affiliation{State Key Laboratory for Modern Optical
Instrumentation, Centre for Optical and Electromagnetic Research,\\
Joint Research Center of Photonics of the Royal Institute of
Technology (Sweden) and Zhejiang University, Zhejiang University, Yu-Quan, Hangzhou, 310027, P. R. China \\
Division of Electromagnetic Theory, Alfven Laboratory, Royal
Institute of Technology, S-100 44 Stockholm,Sweden}

\begin{abstract}
A novel open cavity formed by three 60-degree wedges of a photonic
crystal with negative effective index is designed and studied. The
quality factor of the open cavity can be larger than $2000$. The
influence of the interface termination on the resonant frequency
and the quality factor is studied.
\end{abstract}
 \maketitle

A left-handed (LH) material, which exhibits a negative refraction
index due to simultaneously negative permeability and
permittivity, was first introduced by Veselago in 1967
\cite{Veselago1964P1}. Negative refraction has recently attracted
much attention due to its potential impacts and applications (see
e.g. \cite{Shelby2001,Pendry2000P1,Smith2004P1}). A slab of LH
material with $\epsilon= -1$ and $\mu = -1$ can act as a superlens
(with subwavelength focusing) \cite{Pendry2000P1,Chen2004P1}.
Subwavelength focusing can also be achieved with a slab of a
special photonic crystal (PC) \cite{Notomi2000P1,Cubukcu2003P2,
Cubukcu2003P1,Parimi2003P1,Berrier2004P1}. Negative refraction may
occur in two distinct types of PCs \cite{Foteinopoulou2003P1}: (i)
PCs having  negative effective refractive index $n_{eff}$ near the
center of the Brillouin-zone
\cite{Notomi2000P1,Luo2002P2,Ao2004P1,Zhang2004P1}. For this type,
the dot product of the Poynting vector and the wave vector is
negative (like in an LH material), and consequently it may possess
some behaviors similar to those of an LH material (e.g. focusing
slab-lens satisfying Snell's law \cite{Pendry2000P1}). (ii) PCs
with convex equal frequency contours (EFCs) at a corner of the
Brillouin-zone \cite{Luo2002P1,Hu2004P1,Zhang2004P2}. For this
type, the dot product of Poynting vector and the wave vector is
positive, and $n_{eff}$ can not be deduced. Consequently, its
behavior won't be so similar to that of an LH material.

Open cavity \cite{Notomi2000P1}, which relies on the concept of
negative refractive index, is another amazing property of an LH
material. The open cavity suggested by Notomi \cite{Notomi2000P1}
consists of four alternating rectanglar blocks of two materials
with oppositive refractive indices. A simple ray-trace analysis
can show that there exist many closed ray paths (with zero value
of optical path) running across the four interfaces and thus a
kind of an open cavity (with no surrounding reflective wall) is
formed.  The idea of open cavity is based on the cancellation of
the optical path (defined as the integration of the refractive
index over the ray path) and is straightforward. Since a
homogeneous LH material at an optical wavelength has not been
experimentally realized yet and a photonic crystal of negative
refraction in the optical regime is comparatively much easier to
fabricate, it would be very interesting and desirable to realize
(at least by numerical simulation) an open cavity with a photonic
crystal of negative refraction of LH behavior (type (i)). However,
this has not be achieved though Notomi introduced the idea of open
cavity more than 4 years ago \cite{Notomi2000P1}. In the present
Letter we report an open cavity realized with a photonic crystal
of negative refraction.

\begin{figure}
\centerline{\includegraphics[width=3.5in]{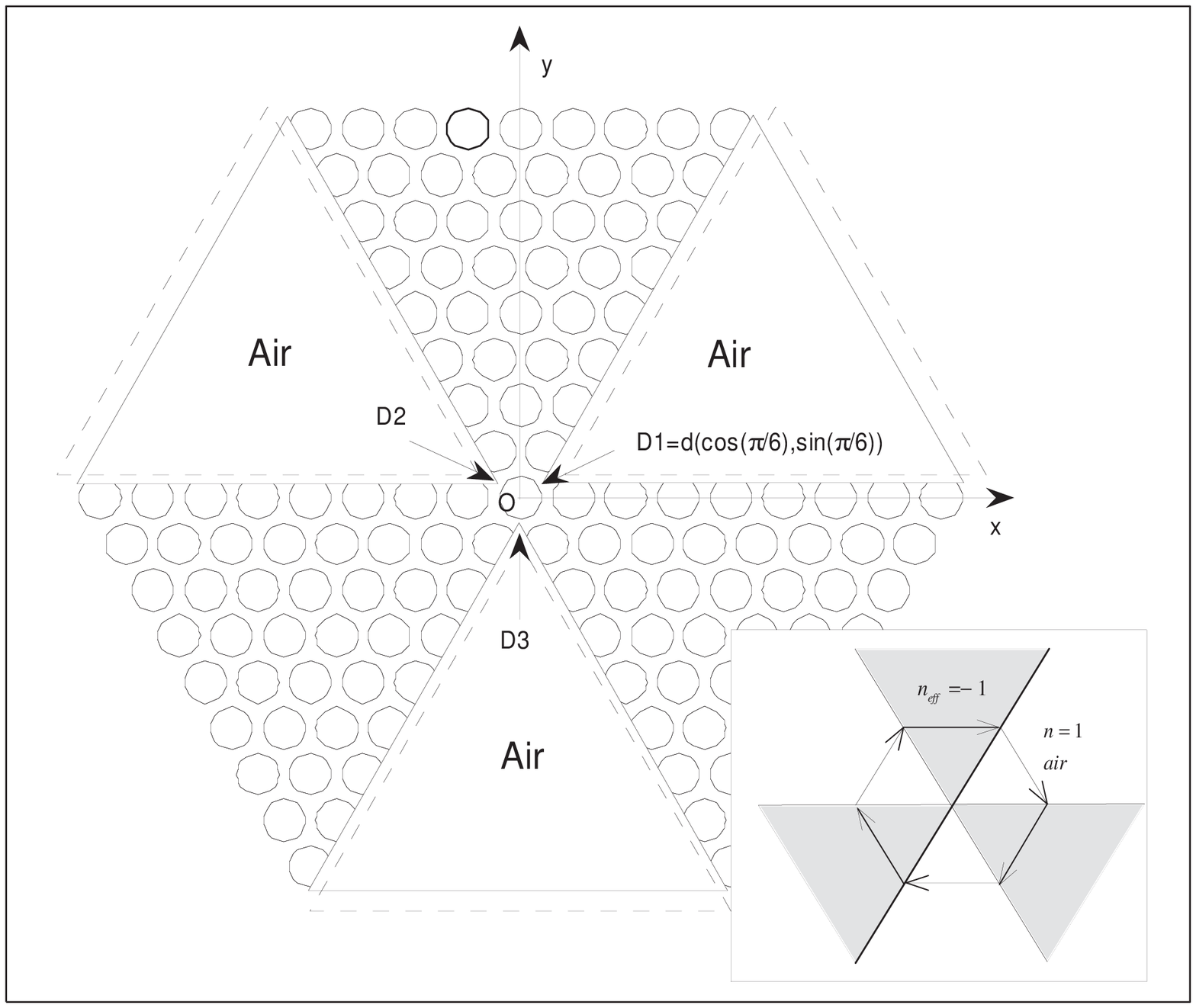}}
\caption{\label{fig:structure} The proposed open cavity formed by
three 60-degree wedges of PC of negative refraction. The position
of surface termination is determined by the distance $d$ between
the center O and any of the tip points. The inset shows a
ray-trace analysis for a closed optical path around the open
cavity.}
\end{figure}

In the design of our open cavity, we use the same 2D photonic
crystal of negative refraction as considered by Notomi
\cite{Notomi2000P1}, i.e., a triangular lattice of air holes (of
radius $0.4a$; $a$ is the lattice constant) in GaAs background
(with $n =3.6$). For E-polarization (with the electric field along
the $z$-direction), the equal-frequency surface is almost circular
(indicating that the PC can be considered as an isotropic medium
and  $n_{eff}$ can be well defined at these frequencies) at the
second band in a frequency window ranging from $ 0.29(c/a)$ to
$0.34(c/a)$. Furthermore, EFS for a higher frequency has a shorter
radius. This indicates that the Poynting vector (group velocity)
is in the opposite direction of the wave vector and thus the
negative refraction is of LH behavior.  In particular, $n_{eff}$
is nearly $-1$ at frequency $f=0.30(c/a)$.

\begin{figure}
\centerline{\includegraphics[width=4.5in]{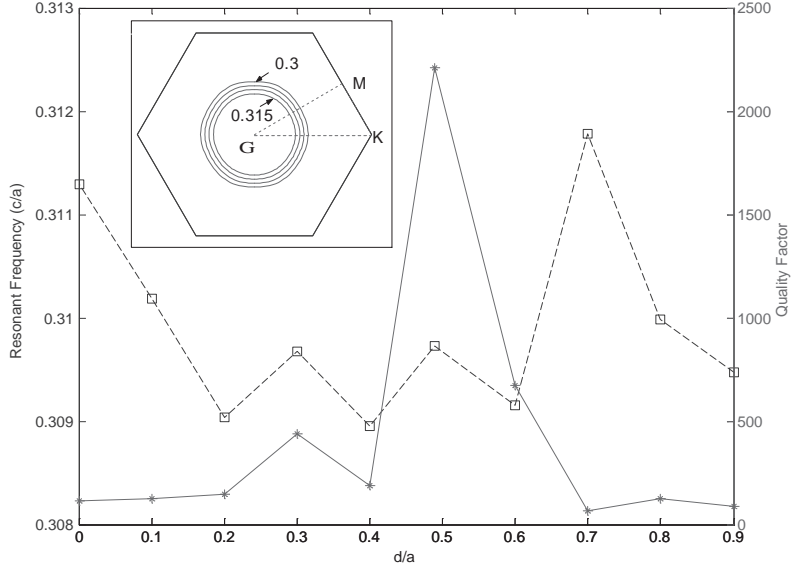}}
\caption{\label{fig:scand} The resonant frequency (squares
connected with dashed lines) and the quality factor (stars
connected with solid lines) when the distance $d$ between the
origin O and the tip of an air wedge increases. Each PC wedge has
13 rows of air holes.  The inset shows the equal frequency surface
for the PC of negative refraction around the resonant frequency
(the numbers mark the frequencies  in  unit  of (c/a)). }
\end{figure}

It was shown in our previous work \cite{Ruan2005P2} that the
direction of surface termination is critical for a high
transmission (i.e., coupling) at  an interface between air $(n=1)$
and a photonic crystal with effective refractive index
$n_{eff}=-1$. In particular, when wave vector $\bf{k}$ is along
the direction  $\Gamma$-M or $\Gamma$-K, the modes have asymmetry
of $C_{1h}$, which gives two distinct cases: the bulk mode has an
even symmetry when $\bf{k}$ is along the $\Gamma$-M direction, and
an odd symmetry when $\bf{k}$ is along the $\Gamma$-K direction.
Since the external plane wave at normal incidence is of even
symmetry, only the Bloch waves with an even symmetry can be
excited. Consequently this results in  small transmission (i.e.,
large reflection) for any incident angle at an interface normal to
the $\Gamma$-K direction and the reflection is small for virtually
all incident angles when the interface is normal to the $\Gamma$-M
direction. Therefore, the open cavity sketched by Notomi won't
work due to the large reflection loss at air-PC interfaces (if one
tries to reduce the reflection at one air-PC interface of a
rectangular block by aligning the normal of this interface with
the $\Gamma$-M direction, the reflection would be large at the
other air-PC interface of the rectangular block).

\begin{figure}
\centerline{\includegraphics[width=2.5in]{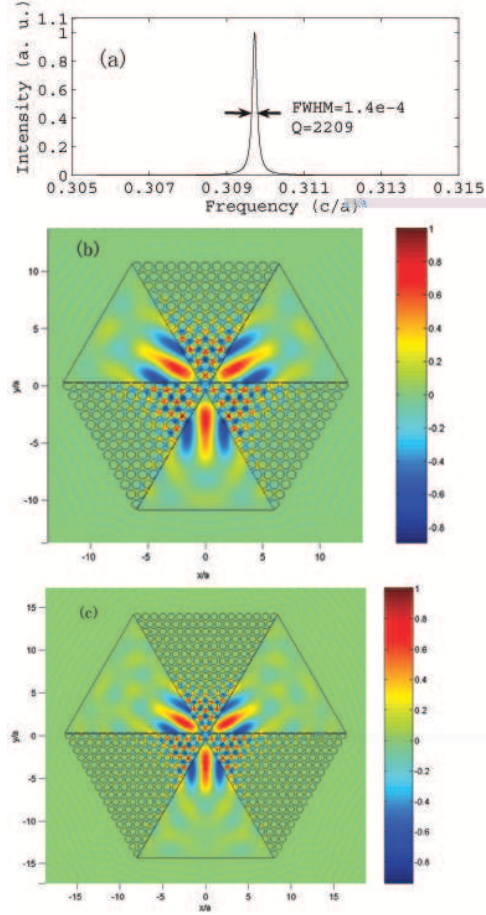}}
\caption{\label{fig:resonant}(a) The spectral response and  (b)
the resonant modal field of the open cavity at resonant frequency
$f= 0.309657(c/a)$. Each PC wedge contains 13 rows of air holes.
The distance between the center O and any of the tip points is
$d=0.49a$.  (c) The resonant modal field of the open cavity with
17 rows of air holes in each wedge.}
\end{figure}

Figure \ref{fig:structure} shows the novel design of our open
cavity based on the same photonic crystal of negative refraction.
It consists of three 60-degree wedges of PC of negative refraction
with three 60-degree in-between wedges of air. The whole structure
is designed in such a way that all the air-PC interfaces are
normal to the $\Gamma$-M direction (to reduce the reflection at
the air-PC interfaces). The simple ray-trace sketch for a close
optical path in the inset of Fig. 1 shows how such an open cavity
may work (if the reflection at the interfaces is small).

Besides the direction of the surface termination, the position of
the surface termination also has a significant influence on the
refraction at an air-PC interface \cite{Xiao2004P1}. In our
design, the position of the surface termination can be determined
by the location of the tip points $D_{1}$, $D_{2}$ and $D_{3}$
(see Fig. \ref{fig:structure}). Due to the symmetry, the
coordinates of these tip points can be determined uniquely by a
single parameter $d$ (the distance between the center O and any of
the tip points), i.e. $D_{i}=d(\cos(i2{\pi}/3-{\pi}/2),
\sin(i2{\pi}/3-{\pi}/2))$, $i=1,2,3$.  In Fig.
\ref{fig:structure}, the dashed lines indicate the position of
surface termination for another value of $d$ (i.e., when the air
wedges move a bit outward).

Finite difference time domain (FDTD) method  \cite{TafloveFDTD} is
used here to calculate the resonant mode and analyze the resonant
property of the present open cavity. The fast Fourier transform
(FFT)/Pad\'{e} approximation \cite{Dey1998P1} is also applied to
compute efficiently the resonant frequency $f_{0}$ and the quality
factor $Q=f_{0}/\Delta f$, where $\Delta f$ is the full width half
maximum (FWHM) of the spectral response of the cavity
\cite{Dey1998P1}. We first study how the position of the surface
termination influences the quality factor of an open cavity with
13 rows of air holes in each PC wedge. Figure~\ref{fig:scand}
shows how the resonant frequency (squares connected with dashed
lines) and the quality factor (stars connected with solid lines)
of the open cavity vary with the distance $d$ between the origin O
and the tip of an air wedge. $d=0.9a$ corresponds to a tangential
cut of the boundary air holes (i.e., they just remain whole). When
$d$ increases from 0 to $0.9a$, the resonant frequency  varies in
a range of $[0.308 (c/a), 0.313(c/a)]$, in which EFS is quite
circular (see Fig.~\ref{fig:scand}). This also indicates that the
resonance can occur only in a frequency range where an effective
negative reflection index can be deduced and is close to $-1$.

From Fig. \ref{fig:scand} (for the case when each wedge has 13
rows of air holes) one sees that the quality factor reaches a peak
value of about 2209 at $d=0.49a$. The spectral response of the
open cavity is plotted in Fig. \ref{fig:resonant}(a) (calculated
with FDTD by illuminating the cavity with a pulse and monitoring
the evolution of the field; the FFT/Pad\'{e} method is applied to
determine the  resonant frequency). From this figure one sees that
the resonant frequency $f_{0}$ is $ 0.309725 (c/a)$ and the FWHM
of the spectral response is $\Delta f=1.4*10^{-4} (c/a)$. Thus,
the corresponding quality factor is $Q \equiv f_{0}/\Delta f
\approx 2209$. Once the resonant frequency is known, we can
determine the corresponding modal field by simulating a second
propagation and extracting the field pattern at that frequency.
Fig. \ref{fig:resonant}(b) shows the corresponding resonant modal
field. The present cavity has a symmetry of point group $C_{3v}$.
From Fig. \ref{fig:resonant}(b) one can see that the irreducible
representation of the localized modes in the open cavity is
$A_{1}$ for point group  $C_{3v}$ \cite{Sakoda}. The modal field
is localized both in the photonic crystal and air. The modal field
distribution has a maximal amplitude in air. From Fig.
\ref{fig:resonant}(b) one can also see obvious negative refraction
at each PC/air interface.

The resonant frequency varies little (within $2*10^{-4}(c/a)$) as
$N_{row}$ increases. This is understandable since the modal field
of the open cavity is localized near the center and consequently
the resonant frequency of the open cavity is determined mainly by
the distribution of the effective refractive index along a closed
optical path near the center. One can easily see that the modal
field distribution in Fig. \ref{fig:resonant}(c)  (with 17 rows of
air holes in each PC wedge) is indeed similar to that in Fig.
\ref{fig:resonant}(b) (with 13 rows of air holes in each PC
wedge). The quality factor increases a bit from 2209 to 2540.

In summary, we have studied a novel open cavity formed by a
photonic crystal with negative effective refraction index. Due to
the negative refraction (and the high transmission) at the PC/air
interfaces, the propagating wave in the air wedges is connected to
that in the PC wedges and thus form a closed path (with zero
optical path due to the negative value of the effective refractive
index of the PC) for resonance. The quality factor of the open
cavity can be higher than $2000$. The influence of the interface
termination on the resonant frequency and the quality factor has
been addressed. The present design idea for an open cavity can be
generalized to other PCs of negative refraction (of type (i)).

\begin{acknowledgments}
This work was supported by the National Basic Research Program
(973) of China (2004CB719800) and the National Natural Science
Foundation of China (under key project No. 90101024 and project
No. 60378037).S. L. He's email address is: sailing@ieee.org
\end{acknowledgments}

\end{document}